\documentclass[prb,twocolumn,citeautoscript,amsfonts]{revtex4}
\usepackage{dcolumn}
\usepackage{epsfig}

\def\TiC{Ti$_8$C$_{12}~$}
\def\C3v{C$_{3v}$}
\def\D2d{D$_{2d}$}
\def\Td{T$_{d}~$}
\def\Th{T$_{h}~$}
\def\cm1{cm$^{-1}~$}
 
\begin{document}
\title{Stability, Electronic Structure and Vibrational Modes of
\TiC  Dimer }
\author{Tunna Baruah$^{1,2}$} 
\author{Mark R. Pederson$^2$}
\affiliation{$^1$Department of Physics, Georgetown University,
Washington, DC 20057, USA}
\affiliation{$^2$Center for Computational Materials Science,
Naval Research Laboratory, Washington, DC 20375-5000, USA}
\date{\today}
 
\begin{abstract}
 We present our density functional results of the geometry, electronic
structure and dissociation energy of \TiC dimer. We show that as 
opposed to the currently held view that \TiC are highly
stable monodispersed clusters, the neutral \TiC clusters form covalent bonds and
form  stable dimers. 
We determine that the Ti atoms bond weakly (0.9 eV/bond) to
organic ligands such as ammonia. Alternatively the Met-Car
dimer has a cohesive energy of 4.84 eV or approximately
1.2 eV per bond. While Met-Car dimers are stable, formation of
these dimers may be quenched in an environment that contains
a significant population of organic ligands. The ionization
and dissociation energies of the dimer are of same order
which prevents the observation of the dimer in the  ion
mass spectroscopy.
The analysis of the vibrational frequencies show the lowest-energy
structure to be dynamically stable. We also present infrared 
absorption and Raman scattering spectra of the \TiC dimer.
\end{abstract}
\maketitle

 The \TiC  clusters consisting of eight metal and twelve carbon
atoms were found to be highly stable \cite{Guo1,Guo2}. Such
clusters of other sizes were found to be less abundant as seen from
the mass-abundance spectrum reported by Castleman and his co-workers
 \cite{Guo1}. 
Apart from Ti atoms, similar clusters with eight metal atoms  were
found to form with other transition
metals such as V, Zr, Hf, Nb, Mo, Fe, Cr \cite{Guo1,Guo2,Wei,Pilgrim1} 
etc. 
These
metallocarbohedrene clusters are  generally refered to as Met-Cars.
A detailed account of the
various Met-Cars can be found in Ref. \cite{Rohmer}.

The ground state structure of Met-Cars has generated a lot of debate.
The early studies have assigned a structure with \Th
symmetry  which later on was shown to be Jahn-Teller unstable \cite{Dance1,Dance2}.
Dance \cite{Dance2} has shown the existence of a barrierless
transition  path from \Th leading to a \Td structure.
Another calculation by Chen {\it et al.}  \cite{Chen} have proposed a 
\D2d structure for the \TiC.  Recently,  Gueorguiev and Pacheco 
\cite{Pacheco}
have shown that Jahn-Teller distorted \Td structure is lowest in energy
among the \Th, \Td, and \D2d structures. A recent calculation by the 
present authors found 
the Met-Cars to possess a \C3v structure which is lower
in energy than either the \Td or the \D2d structures \cite{metcar1}. 

An important aspect of these clusters is their  paradoxical
nature - while they are highly stable in vacuum they are found to be
very air sensitive.  The ionization potential and 
electron affinity of the Met-Cars suggest that these clusters have
moderate reactivity \cite{Duncan,Sakurai,Wang2,metcar1}. 
  Another such issue which  has  not yet  been studied is the  
formation of larger clusters in which the individual Met-Cars are 
the building blocks.  It may be mentioned here that Met-Cars of
various sizes, e.g. T$_{13}$C$_{22}$ exist  but their 
structures show them to be different from \TiC \cite{Wang1,Wang3}.
They
can not be considered as made up from Met-Car units. In this article, 
we examine the possibility of two Met-Cars
coalescing to form a stable dimer. We show that a  dimer connected
through  four-membered rings  of Ti and C atoms form a stable
structure in which both the individual Met-Cars retain their
identity. This may form a basis for future studies on Met-Car based
molecular solids. We also present the calculated infrared absorption and
the Raman spectra of the \TiC dimer. We believe this may help
in identifying the Met-Car dimers in experimental IR and Raman
spectra.

The theoretical calculations were carried out within the density
functional theory \cite{Hohenberg} using a linear combination of
atomic-orbitals  (LCAO) approach.  The calculations were
performed using the NRLMOL \cite{NRLMOL} program which employs 
a Gaussian basis set.
The molecular orbitals were expanded as linear combinations of
Gaussian functions centered at the atomic sites.
The exponents for the single Gaussians have been fully optimized
for DFT calculations \cite{pore99}.  
The calculations were carried out at the all-electron level
with the generalized gradient approximation to the
exchange-correlation functional \cite{PBE}. A variational mesh
is used to reduce the computation as well as storage of data and
still maintain the high accuracy of the calculations \cite{mesh}.
 The code is massively 
parallelized  \cite{pss2000} which allows one to perform calculations on a cluster of PCs.
The  geometry optimization was carried out using the conjugate-gradient
technique. The self-consistent calculations for each geometry was
carried out till the energy difference fell below 
$1.0 \times 10^{-6}$ a.u., so as to obtain accurate forces. The force minimization was carried out to a tolerance limit 
of 0.001 a.u on each atom. 

The calculation of infrared absorption and Raman scattering spectra were
performed according to the method described in Ref. \cite{Porezag1996}.
The dynamical matrix is generated by  moving the atoms by $\sim \pm$  0.08 a.u. from their equilibrium positions. The Raman and infrared 
intensities are obtained by applying an electric field of strength 
0.005 a.u. and calculating the first and second derivatives of the
dipole moment and polarizability with respect to the applied electric
field.

\begin{figure}
\epsfig{file=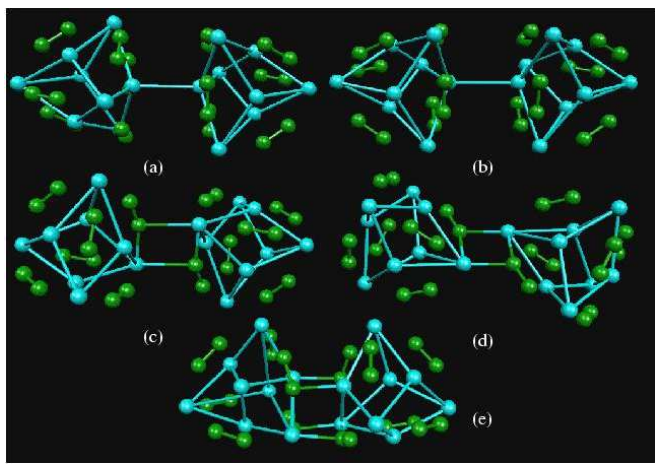,width=\linewidth,clip=true}
\caption{ The  various dimer configurations considered in the
present work. The metal cage structure  with the carbon dimer on the
surface of each Met-Car is highlighted.  Also, the bonds between the 
Met-Cars are shown. 
}
\label{fig1}
\end{figure}

  Our earlier calculations have shown that the  lowest energy 
structure of the neutral \TiC  possesses \C3v symmetry \cite{metcar1} 
while the \D2d and the \Td  structures lie slightly higher in
energy. We have also determined that both the \C3v and \D2d 
structures are vibrationally stable. 
To determine the lowest-energy structure of the \TiC dimer, we 
started out by considering several different dimers composed
of two \C3v structures.   A schematic diagram of the optimized 
structures of the various dimers is shown in Fig.  \ref{fig1}. 
The figure shows the metal cages with the
carbon dimers lying on the surfaces of each Met-Car. The bonds
between the Met-Cars are also pointed out. 
The first structure that we studied retains
the \C3v structure and tests whether a  Ti-Ti ``bond'' may form. This
corresponds to structure (a) in Fig.~\ref{fig1}.
Since the two Met-Cars are related by inversion, the overall
symmetry of this dimer is D$_{3d}$.  This structure may exist in 
either the staggered or eclipsed form (Fig. \ref{fig1}(b)). The
second structure that we considered was determined by allowing
two C-C dimers to form 4-membered rings. This corresponds to 
structure (c) in Fig.~\ref{fig1}.  The underlying C$_{3v}$
symmetry of this structure is broken and the final symmetry consists
of the 8-membered group generated by all reflections.  Another 
structure, labeled (d) in the figure, may be generated by allowing 
4-membered Ti-C-Ti-C rings to connect the two Met-Cars. This 
structure is bound relative to two C$_{3v}$ Met-Cars by 2.9 eV. The 
four-membered ring is composed of intra- and inter- molecular Ti-C 
bonds of 1.96 \AA~ and 2.15  \AA~ respectively.
The symmetry group is generated by inversion and (xyz)$\Rightarrow$
(zyx). Examination
of this structure shows that an intermolecular Ti$_4$C$_4$ cubane unit 
may be formed by bending the respective Met-Cars out of the planar ring
and toward one another. This leads to structure (e) in Fig.~\ref{fig1}.
The resulting TiC is composed of bondlengths that average 2.18 \AA~.  
However, the cubane connector is distorted with both long (2.32 \AA~) 
and short (2.04 \AA~) bonds forming the intramolecular faces and medium
(2.18 \AA) bonds forming the intermolecular bonds. 
 This dimer structure is comprised of  4 Ti-C
  bonds between the neighboring balls. It is bound
  by 4.84 eV or approximately 1.21 eV per Ti-C
  bond. This structure has only one symmetry operation
  (xyz) $\Rightarrow$ (-x,y,-z).  The binding energy per Ti-C
  bond is similar to the energy we found for the
  Ti-N bonds in our titration calculations discussed
  later.
Within the symmetry constraints discussed above, the geometry of these 
structures were modified until the forces on atoms
were less than 0.001 a.u. We wish to point out that while carrying
out the optimization, the symmetry of the individual Met-Cars was
not enforced, rather the Met-Cars were allowed to relax while
forming a dimer. This will allow the individual MetCars
to relax into symmetry allowed structures other than \C3v while 
forming the dimer.

We find the structure
labelled as (e) to possess the lowest energy  among all the structures
considered.  In Table \ref{table1}, we present the relative energies
of the dimers with respect to the energy of (e). The dissociation
energies required for the dimers to dissociate into the constituting
Met-Cars are also presented. The dissociation energy of the structure
(e) is highest showing it to be the most structurally stable form.
In Table \ref{table1} we also present the HOMO-LUMO gaps of all
the structures.
The HOMO-LUMO gap of the (b) structure is highest 
followed by that of the structure (e) while in structures (c) and (d)
the gap is small which will show more metallic behavior and also
higher chemical reactivity. To assess the chemical reactivity of the
lowest-energy dimer, we also calculated the vertical ionization
potential and electron affinity(VEA). In this calculation, the
 charged clusters were not allowed to relax.  The VIP of the structure (e) is
4.4  eV while the VEA is 1.3 eV. The VIP is of the same order
as that of a single Met-Car (4.47 eV) while the VEA of the dimer is
higher than that of \TiC (1.05 eV) \cite{metcar1}. The reactivity
of the dimer is therefore higher than that of single \TiC. 

\begin{figure}[h]
\epsfig{file=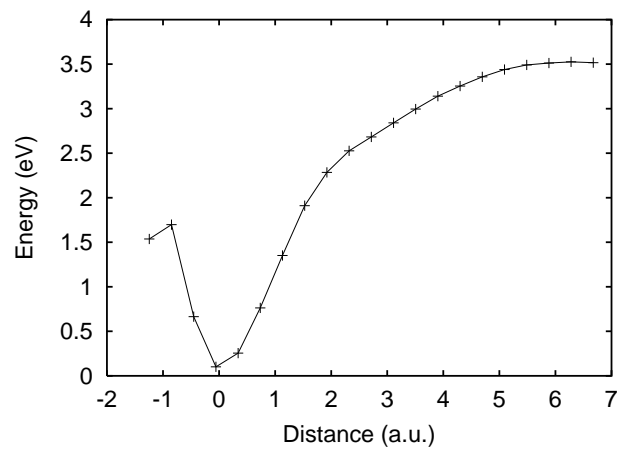,width=\linewidth,clip=true}
\caption{ The potential well formed when two Met-Cars
are brought closer without relaxation. The relative
energies and the distances are given with respect to
the dimer (d) in Fig. \ref{fig1}.
}
\label{Evsr}
\end{figure}

To investigate the retention of individual symmetry of the Met-Cars 
while forming a dimer, we studied the energy surface of the dimer as 
a function of \TiC-\TiC distance. In this case, the symmetry of the 
Met-Cars were maintained while the distance between them was varied. 
The energy vs. dimer separation curve is shown in Fig. \ref{Evsr}. 
The minimum in the curve indicates the
equilibrium separation when the individual Met-Car's geometries
do not relax.  This minimum energy point is close to the structure 
(d) in Fig. \ref{fig1}.  The difference between the equilibrium 
point in Fig. \ref{Evsr} and the structure (d)
arises from the fact that in structure(d) the symmetry of the individual
Met-Cars were not retained and therefore the energy of structure (d) is
lower. The small energy difference (0.1 eV) suggests that the  
symmetry distortion is small in the structure (d).

  Another relevant question is how stable a molecular solid made from
\TiC would be.  Our calculations show that an isolated Met-Car dimer 
would form chemical bonds and lead to a binding energy per Met-Car of
approximately 2.42 eV. For chemically bound systems, this is on the 
weak side which
  suggests that a condensed phase of Met-Cars may behave differently.
  For example, when the Met-Car center-to-center distance is increased
  by 2.0 au, the intermolecular chemical bonds are completely broken and the
  isolated structure of the Met-Cars is retained. However we
  find that this structure is still bound by 0.285 eV/Metcar which
  would suggest an FCC lattice of Met-Cars would be bound by roughtly
  12*0.285/2 = 1.8 eV per Met-Car. Such a lattice might be further
  stabilized by some metallic bonding due to the unclosed electronic
  shell. As such the weak covalent bonds formed for a Met-Car
  dimer may not be strong enough to survive in the condensed phase
  and a lattice of isolated Met-Cars is at least competitive.
  The binding energy of the dimers for the same 
particle-particle separation but averaged over different  
orientation
  is about 0.06 eV/Met-Car  which by the above analysis shows
a binding energy in the crystalline environment to be about 0.36 eV.
  This is much more weak than predicted by the earlier considerations.
  Additional calculations on the bulk phase would be useful for
  clarifying this point.

While our results show that the Met-Car dimer is stable with
respect to separated Met-Cars, observation of this dimer will
almost certainly require an environment devoid of ligands.
 The  reactivities of 
Met-Cars have not yet been fully explored.  Some investigations
on the reaction of Met-Cars with molecular oxygen have been reported
\cite{Selvan,Deng,Yeh}. An ab initio calculation by Ge {\it et al.}
\cite{Ge} on \TiC (H$_2$O)$_8$ and \TiC(C$_2$H$_4$)$_4$ have shown
the stability of the \TiC to increase upon reaction with
water while it decreases in reaction with C$_2$H$_4$. Another
calculation by Poblet {\it et al.} \cite{Poblet} have shown an
exothermic reaction for addition of Cl, NH$_3$, CO, C$_6$H$_6$.
They also showed that addition of eight ammonia or Cl is easy
while adducts with eight CO or benzene is difficult.
 To check whether the energy ordering of Met-Cars with different
 symmetry is influenced by 
absorption of radicals, we have carried out optimizations of the 
geometries of the Met-Car in \Td and \Th and eight ammonia molecules 
bonding with the Ti atoms. The energy of the ammonia covered \Td 
structure is significantly lower than that of the \Th structure. 
However, the vertical ionization potential (VIP) of these two structures
do not follow the same trend. The VIP of the ammonia covered \Th
structure is 2.76 eV while that of the \Td structure is 2.41 eV. 
The binding energy of the 
lowest structure with respect to the dissociation into eight ammonia
molecules and the \TiC is 7.43 eV which suggests an average of 0.93 eV 
per bond. The structure of the Met-Cars after absorption of eight 
ammonia retains the symmetry thereby implying that the structure 
does not change upon reaction.
Our calculated Ti-NH$_3$ binding energy of 0.9 eV is similar to
other calculated Ti-ligand binding energies. For example,
Ref. \cite{Ge} and  \cite{Poblet} find binding energies for Cl, 
NH$_3$, CO, and
to H$_2$O be in the range of 0.5 eV (CO) to 2.3 eV (H$_2$O).
In an ammonia rich environment the system
could either form a dimer with at most 12 adsorbed NH$_3$ ligands or two
monomers with a total of 16 adsorbed NH$_3$ ligands. Steric effects
would probably prevent a total of 12 adsorbed NH$_3$ ligands on the
dimer and would also weaken the Ti-NH$_3$ bonds. The former
structure would have energy of at most 15.6 eV 
while the latter structures would have an energy of 14.4 eV.
Steric effects would probably lower the ligandated
dimer below that of two ligandated monomers and kinetic considerations
would also favor growth of the separated ligandated monomers.
These considerations may also explain why the Met-Cars seem to
be stable in soot. 
 
  To facilitate an understanding of the possible molecular processes
which could hinder the observation of the dimer formation, we
have compared  the dimer  total energy with that of Ti$_{13}$C$_{22}$ and  
Ti$_3$C$_2$. The Ti$_{13}$C$_{22}$ is earlier reported to be stable 
\cite{Wang1,Wang2}   which has a cubic structure .
We found that such a process would
cost an energy of 12.3 eV  for the lowest-energy structure 
predicted in 
Ref. \onlinecite{Wang1} which ruled out any possibility of the
dimer fragmentation into these products. On the other hand, 
the ionization energy of the dimer ($\sim$ 4.4 eV) is of the same order as the
energy required for dissociation into monomers (4.84 eV). Since the
mass spectrography measurements are carried out on ionized clusters,
the ionization process is likely to induce dissociation. 
A detailed investigation
of the dimer cation is beyond the scope of the present work.

 We  have also calculated the vibrational modes of the lowest-energy
structure of the dimer. The calculations show
the dimer to possess no imaginary frequencies thereby indicating 
that the structure
(e) indeed refers to an equilibrium structure.  One noticeable
feature of the vibrational modes is that this structure exhibits
a few low frequency modes which are absent in the individual
Met-Car vibrational modes \cite{metcar1}. 

\begin{figure}
\epsfig{file=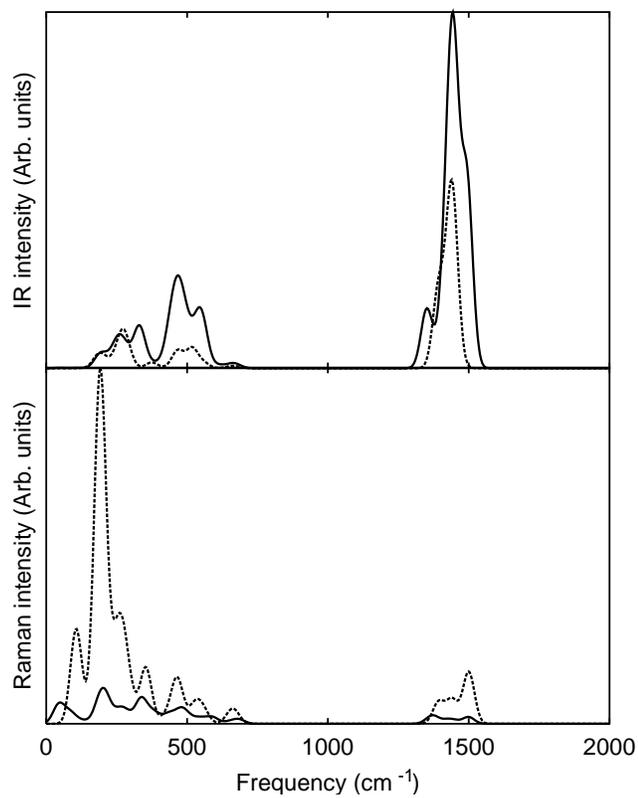,width=\linewidth,clip=true}
\caption{ The  infrared absorption  and Raman scattering spectra 
of the lowest-energy structure of \TiC dimer (solid lines) and the \TiC monomer 
(dashed lines) with \C3v symmetry. 
The intensities are broadened with a Gaussian of FWHM of 40 \cm1.
}
\label{IR}
\end{figure}

The IR absorption spectrum
of the dimer is shown in Fig. \ref{IR} (upper panel). The spectrum has been broadened
with a Gaussian of full width half maximum of 40 \cm1. 
In the low frequency region, peaks occur
at 186, 241, 260, 272, 280, 292, and 330 \cm1 which corresponds to
Ti-C stretching and also twisting modes.  In the single Met-Car, IR active peaks at the
low frequency range are observed at 183, 204, 244, 266, 271, 373, 467, 517, and 559 \cm1 
which are due to Ti-C stretching mode \cite{Heijnsbergen,Pacheco,metcar1}.
In the slightly higher frequency region, the dimer IR spectrum closely resembles that of the
Met-Car IR spectrum with a relatively large peak around 460 \cm1 and
 550 \cm1. These are actually composed of peaks at  460, 464, 
472, 480, 527, 541, 551 \cm1. These correspond to   the C-C twisting
modes and also translation of a carbon dimer on each Met-Car which
leads to stretching of the Ti-C bonds. 
In the high frequency region,
the spectrum shows a 
pronounced broad
peak slightly beyond 1400 \cm1. In fact this broad peak consists
of three narrowly spaced peaks at 1411, 1417, and  1446 and another
 large peak at 1493 \cm1.  These correspond to C-C stretch
modes in which all carbon dimers except the ones forming the four
membered ring, take part.  Other two
smaller peaks appearing at  1349 and 1354 \cm1 are due to the
C-C stretch mode involving only the carbon dimers forming the \TiC-\TiC
bonds. 
It may be mentioned here that our calculation of the
Met-Car IR spectrum shows a pronounced peak near 1400 \cm1 which
also arises due to the C-C stretch mode.  In the dimer, this peak
appears to be shifted to the high frequency region. 


Figure \ref{IR}(lower panel) shows the Raman scattering spectrum of the \TiC dimer.
In this case, a broad peak occurs in the low frequency region below 700
\cm1. As mentioned earlier, there are a few low frequency vibrational
modes which are Raman active. The first peak is comprised of two 
narrow peaks occuring at 44 and 84 \cm1. These correspond to
\TiC-\TiC bond stretch modes.
The highest  peak at about 200 \cm1 consists
of narrow peaks  at
185, 203,  and 207  \cm1 corresponding to twisting and squeezing
of the Ti-C and Ti-Ti inter-Met-Car bonds. The shoulder seen about 
260 \cm1 is also due to Ti-C bonds forming the Ti$_4$C$_4$ cubane 
joing the two Met-Cars. The other noticeable peaks in the broad
spectrum around 337 \cm1 occur due to Ti-C stretching where the
Ti is part of the cubane, around 480 which is due to Ti-C squeezing
mode. The smaller peaks around 591 \cm1 and 687 \cm1 are due to, respectively, 
the C-C stretching
mode where one atom of the  C-C dimer is a member of the cubane and
a rocking motion of some of the carbon dimers which do not
form part of the cubane. The high frequency spectrum shows two 
prominent peaks. One peak consists of narrowly spaced peaks at
1367 and 1375 \cm1 and the other of peaks at 1417, 1448 and 1500 \cm1.
All of these correspond to C-C stretching modes of which the 1367 and
1375 \cm1 correspond to the stretching and squeezing  of the C-C dimers
involved in the formation of the cubane. The above results show that
the  vibrations involving Ti-C bonds occur at lower frequency and are
Raman active whereas the C-C stretch modes occur at high frequecy and
are IR active. Similar features are seen in case of the IR and Raman
spectrum of the single Met-Car \cite{metcar1} although the spectrum is
slightly shifted in frequency. The  low frequency modes ($<$ 100 \cm1)
seen in the
dimer are not visible in the single Met-Car spectra  \cite{metcar1}
 which as mentioned
above arise from the vibrations involving the Ti and C atoms forming
the Met-Car-Met-Car bonds.

In conclusion, we have carried out first principle density
functional based study of the structure and stability  of \TiC dimer.
Our calculations have shown the dimer to be stable by 4.84 eV. An
analysis of the vibrational frequencies show the structure to be
a local minimum. The IR and Raman spectrum of the dimer are presented
and analyzed. The IR active modes are found in the high frequency
region while the Raman active ones are located in the low-frequency
region of the spectra. The symmetry of the individual \TiC 's are somewhat
modified in the formation of the dimer.
Our calculation predicts that a Met-Car solid will be weakly bound.
From the ionization potential and the dissociation energy of the 
dimer, we predict that during the ionization process in the mass
spectrography, the dimer is likely to dissociate into two
Met-Car monomers which accounts for the lack of experimental
observation of the dimer.



TB and MRP were supported in part by ONR grant N0001400WX2011.


\begin{table}
\caption{ The energies of the Met-Car dimers relative to the lowest 
energy structure, the dissociation energies (E$_d$) and the HOMO-LUMO
gap ($\Delta$) of the [Ti$_8$C$_{12}$]$_2$ clusters. The labels of the
clusters are consistent with the Fig. \ref{fig1}. }

\begin{tabular}[t]{lccc} \hline
 
& Relative & Dissociation &   HOMO-LUMO      \\
 
& energy &  energy  &  gap\\
         & eV     & eV   &   eV   \\ \hline
 
(a)   &  4.17  & 0.67       & 0.08  \\
(b)   &  4.17  & 0.67       & 0.22  \\
(c)   &  3.66  & 1.18       & 0.03  \\
(d)   &  1.98  & 2.86       & 0.04  \\
(e)   &  0.00  & 4.84       & 0.13  \\  \hline
 
\end{tabular}
\label{table1}
\end{table}


\end{document}